\documentclass[sort&compress,preprint,12pt]{elsarticle}
\usepackage{amssymb}
\usepackage{amsmath}
\usepackage{graphicx}
\usepackage{epstopdf}
\usepackage{geometry}
\usepackage{hyperref}
\hypersetup{colorlinks=true,%
            citecolor=blue,%
            filecolor=blue,%
            linkcolor=blue,%
            urlcolor=blue,%
            pdftex}

\journal{xxx}

\begin{document}
\begin{frontmatter}

\title{{\bf{Finite-size scaling of geometric renormalization flows in complex networks}}}
\author[label1]{Dan Chen}
\author[label1]{Housheng Su\corref{cor}}
\author[label2]{Xiaofan Wang}
\author[label3]{Gui-Jun Pan}
\author[label4]{Guanrong Chen}
\cortext[cor]{Corresponding author: shs@hust.edu.cn}
\address[label1]{School of Artificial Intelligence and Automation, Image Processing and Intelligent Control Key Laboratory of Education Ministry of China, Huazhong University of Science and Technology, Wuhan, 430074, China}
\address[label2]{Department of Automation, Shanghai University, Shanghai 200072, China}
\address[label3]{Faculty of Physics and Electronic Science, Hubei University, Wuhan 430062, China}
\address[label4]{Department of Electrical Engineering, City University of Hong Kong, Hong Kong, China}

\begin{abstract}
Recently, the concept of geometric renormalization group provides a good approach for studying the structural symmetry and functional invariance of complex networks. Along this line, we systematically investigate the finite-size scaling of structural and dynamical observables in geometric renormalization flows of synthetic and real evolutionary networks. Our results show that these observables can be well characterized by a certain scaling function. Specifically, we show that the critical exponent implied by the scaling function is independent of these observables but only depends on the small-world properties of the network, namely, all networks located in the small-world phase have a uniform scaling exponent, while those located in the non-small-world phase and in their critical regions have another uniform scaling. More importantly, we perform extensive experiments on real evolutionary networks with small-world characteristics, and our results show that these observables also have uniform scaling in their geometric renormalization flows. Therefore, in a sense this exponent can be used as an effective measure for classifying universal small-world and non-small-world network models.
\end{abstract}


\end{frontmatter}

\section*{Introduction}
Complex networks have attracted considerable attention from various disciplines such as mathematics, physics, biology, engineering, computer science, and so on~\cite{Boccaletti2006}. Its emergence was due to the discovery that associated with a real system there exists a corresponding network which can well define the interactions among the system components~\cite{Barabasi2016}. Because of this, one can better understand the structural and functional properties of the real system. In particular, by studying the topology of the network associated with a real system, some small-world properties~\cite{Watts1998} and scale-free properties~\cite{Barabasi1999} were identified. Consequently, their functional performances could be further studied, including synchronization~\cite{Arenas2008,Nicosia2013}, observability and controllability~\cite{Liu2011,Liu2013}, reaction-diffusion~\cite{Colizza2007}, navigation~\cite{Kleinberg2000}, transportation~\cite{Lopez2005,Li2010}, and many other dynamic behaviors.

However, exploring the network structural and functional properties also faces many challenges. For example, the evolutions of real-world networks usually lead to more complicated interactions and increasing numbers of nodes and edges, causing more difficulties to the investigations. In the past two decades, the renormalization technique ~\cite{Newman1999,Kadanoff2000} was found to be very effective for tackling the troublesome problem. This framework significantly reduces the size and the complexity of a large-scale network by retaining the ``slow" degrees of freedom in the network, while integrating the rest together. In so doing, smaller networks can be used to approximate the initially large ones. By performing a coarse-graining procedure for a spatially embedded scale-free network, it was shown~\cite{Kim2004} that a smaller network can maintain important structural characteristics of the original one. Based on random walks, a coarse-graining method was proposed~\cite{Gfeller2007} to reduce the size of a network, but retain most spectral properties of the original network through an iteration process. Furthermore, using the shortest path-length measure of a network, a box-covering technique was presented~\cite{Song2005,Song2006,Song2007} for reducing the size of the network. It was shown that the iteration process can keep an approximately identical degree distribution of the network, which was verified by some real-world networks. Such a property is known as the network self-similarity~\cite{Song2005}. In the following years, the box-covering technique had significant influences on the research of fractality and self-similarity~\cite{Goh2006,Kim2007a,Kim2007b}, as well as flows and fixed points~\cite{Radicchi2008,Radicchi2009,Rozenfeld2010}, of various complex networks.

It was observed that, for networks with small-world or even ultrasmall-world properties, the transformation method based on shortest path-length cannot be effectively applied to study their structural symmetry and functional invariance. Therefore, a geometric renormalization (GR) framework was proposed~\cite{Garcia2018}, embedded in a hidden metric space~\cite{Serrano2008,Krioukov2009,Krioukov2010,Papadopoulos2012,Boguna2010,Allard2017}, which provides deep insights for studying the structural symmetry of complex networks. This framework is proved capable of preserving both structural and dynamical characteristics of scale-free networks, such as degree distribution, clustering spectrum, dynamics, and navigability, to a certain extent of accuracy within an appropriate number of iterations. Due to the finite-size effect on such networks, however, excessive GR iterations will eventually result in a large deviation of the network features from the original ones.

To further explore the variation of the characteristics of a network in the GR iteration process, this paper reports a comprehensive study of the finite-size scaling (FSS) behavior of the structural and dynamical observables in renormalization flows of synthetic and real evolutionary networks. Specifically, we consider networks in small-world phase, non-small-world phase, and their critical regions, respectively, and perform FSS analyses of their structural and dynamic observables. Our results show that these observables can be characterized by a certain scaling function, and the exponent implied by the scaling function is independent of these observables but only depends on the characteristics of the small-world structure of the network. We have verified the effectiveness of this conclusion via some real evolutionary networks, which further provides evidence for the predictive power of synthetic models to real systems.

\section*{Geometric renormalization group and observables of the network}
We carry out a comprehensive study of the finite-size scaling (FSS) behavior of the structural and dynamical observables in renormalization flows of the typical $\mathbb{S}^1$ geometric network model~\cite{Serrano2008}. This model is generated by two mechanisms of popularity and similarity dimensions~\cite{Papadopoulos2012}, which can clearly explain some universal properties of real networks, regarding their structural complexity, evolutionary mechanism, and dynamic behavior. First, an $\mathbb{S}^1$ network is generated with $N_0$ nodes and $E_0$ edges, denoted as $G_0$. Then, starting from $G_0$, non-overlapping blocks of continuous nodes of size $s$ are defined in the $\mathbb{S}^1$ circle~\cite{Garcia2018}. Here, $s = 2$ is chosen, and one-step GR iteration is performed to obtain $G_1$. Consecutively, a layer-$l$ renormalized network $G_l$ is obtained after $l$ steps of GR iterations. The number of nodes in layer $l$ is denoted by $N_l$, and the relative network size of layer $l$ is $n_l = N_l/N_0$. Recall that the GR transformation has a good characteristic that it can well predict the average degree of a large-scale network in the process of GR iterations. Taking the $\mathbb{S}^1$ model for example, it was shown~\cite{Garcia2018} that the average degree can satisfactorily approximate the exponential relation through the GR flows, with $\langle k\rangle_{l}=s^{\alpha}\langle k\rangle_{l-1}$, where $\langle k\rangle_l$ is the average degree of the renormalized network $G_l$, and the exponent $\alpha$ depends on the structural parameters $\nu$ and $\sigma$ of the $\mathbb{S}^1$ model (see Supplementary Information for notations and descriptions). With respect to different phases of structural and dynamical behaviors, the $\mathbb{S}^1$ model can be roughly divided into three regions~\cite{Garcia2018} (see section II in the Supplementary Information), denoted as \textbf{I}, \textbf{II}, and \textbf{III}, respectively. In regions \textbf{I} and \textbf{III}, which correspond to the small-world phase, $\langle k\rangle_{l}$ approximates an exponential growth. In region \textbf{II}, which corresponds to the non-small-world phase, $\langle k\rangle_{l}$ approximates an exponential decay, where the structure of the renormalized network is similar to a ring as a fixed point. At the edge of the transition between regions \textbf{I} and \textbf{II}, $\langle k\rangle_{l}$ presents a tendency of slow increase with the increase of $l$, which gradually becomes saturated (see Supplementary Information Fig.~S1 for details).

In the following, a GR flow is tracked by eight observables. One is the normalized maximum degree of the renormalized network,
\begin{equation}\label{Eq:1}
k_{l, \max }=\frac{K_{l}}{N_{l}-1},
\end{equation}
where $K_l$ is the maximum degree of the renormalized network $G_l$, and $N_l - 1$ is the maximum value that $K_l$ may take. Another observable is the normalized average degree,
\begin{equation}\label{Eq:2}
\langle k\rangle_{l, n}=\frac{\langle k\rangle_{l}}{N_{l}-1},
\end{equation}
where $N_l - 1$ is the maximum value that $\langle k \rangle_l$ may take. The average clustering coefficient $\langle c \rangle_l$ and the average shortest path-length $\langle \ell \rangle_l$ for the renormalized network $G_l$ will also be considered below.

\begin{figure}[!h]
\centering
\includegraphics[angle=0,width=17.5cm,height=14cm]{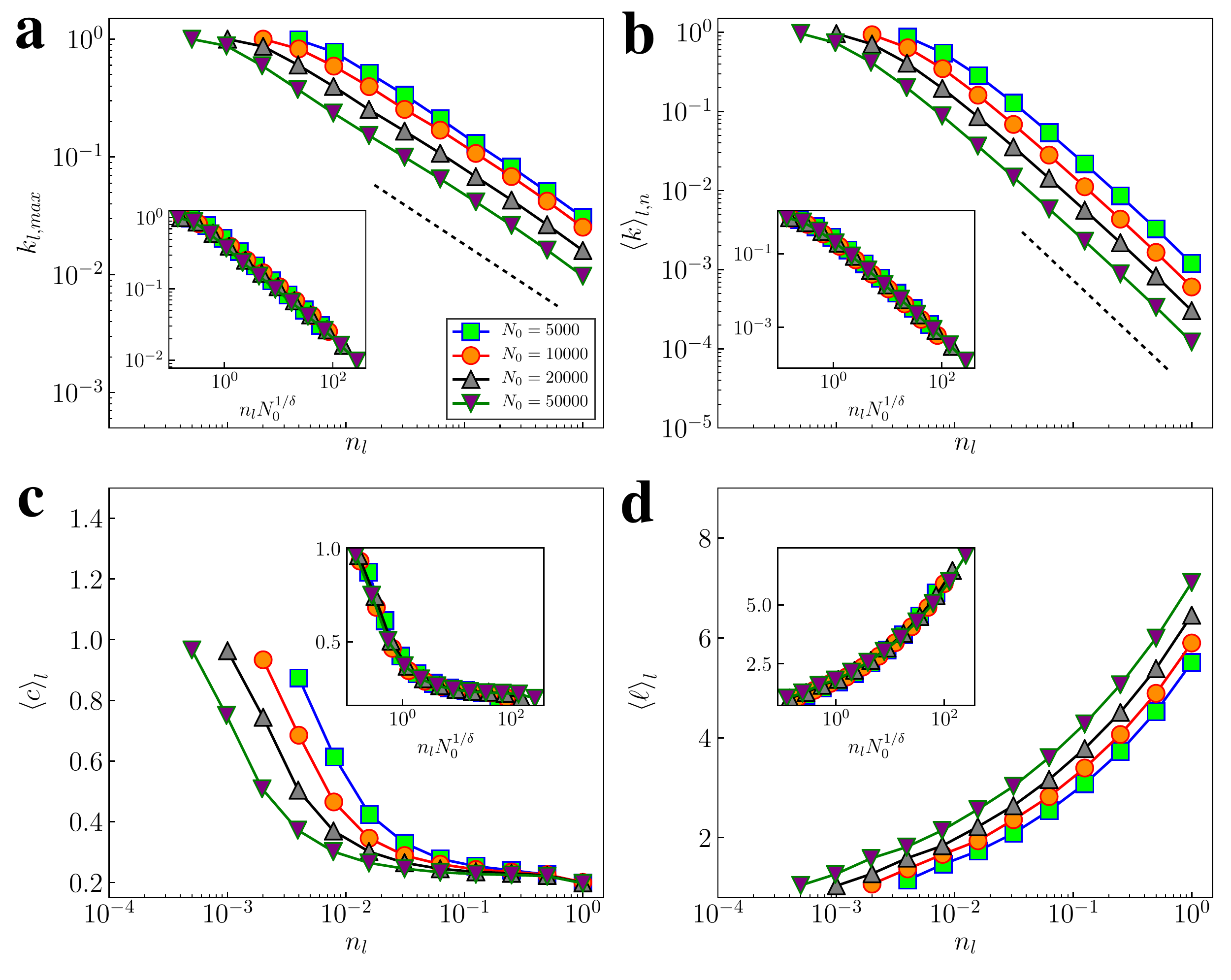}
\caption {FSS analysis of $\mathbb{S}^1$-network topological observables along the GR flow. The main figures show each observable as a function of the variable $n_l$ in the process of GR transformation, and the inset shows their scaling functions related to the variable $n_{l} N_{0}^{1/\delta}$. Key parameters of the $\mathbb{S}^1$ network are $\nu = 2.5$ and $\sigma = 1.5$, respectively; the scaling exponent corresponding to four observables is $\delta = 2 \pm 0.1$. For observables $k_{l, max}$ and $\langle k\rangle_{l, n}$, the black dashed-lines predict their power-law behavior. The average hidden degree $\langle\kappa\rangle_0 \approx 6$ and the expected average degree $\langle k \rangle_0 \approx 6$ for all initial networks. All the results are averaged over 10 independent realizations.}
\label{Fig:1}
\end{figure}

Regarding the above characteristics of networks, the FSS behavior is investigated through a box-covering procedure~\cite{Radicchi2008,Radicchi2009}. These observables are commonly used to characterize the basic topological properties of the network, referred to as topological observables of the network, as illustrated in Fig.~\ref{Fig:1}.

\begin{figure}[!h]
\centering
\includegraphics[angle=0,width=17.5cm,height=14cm]{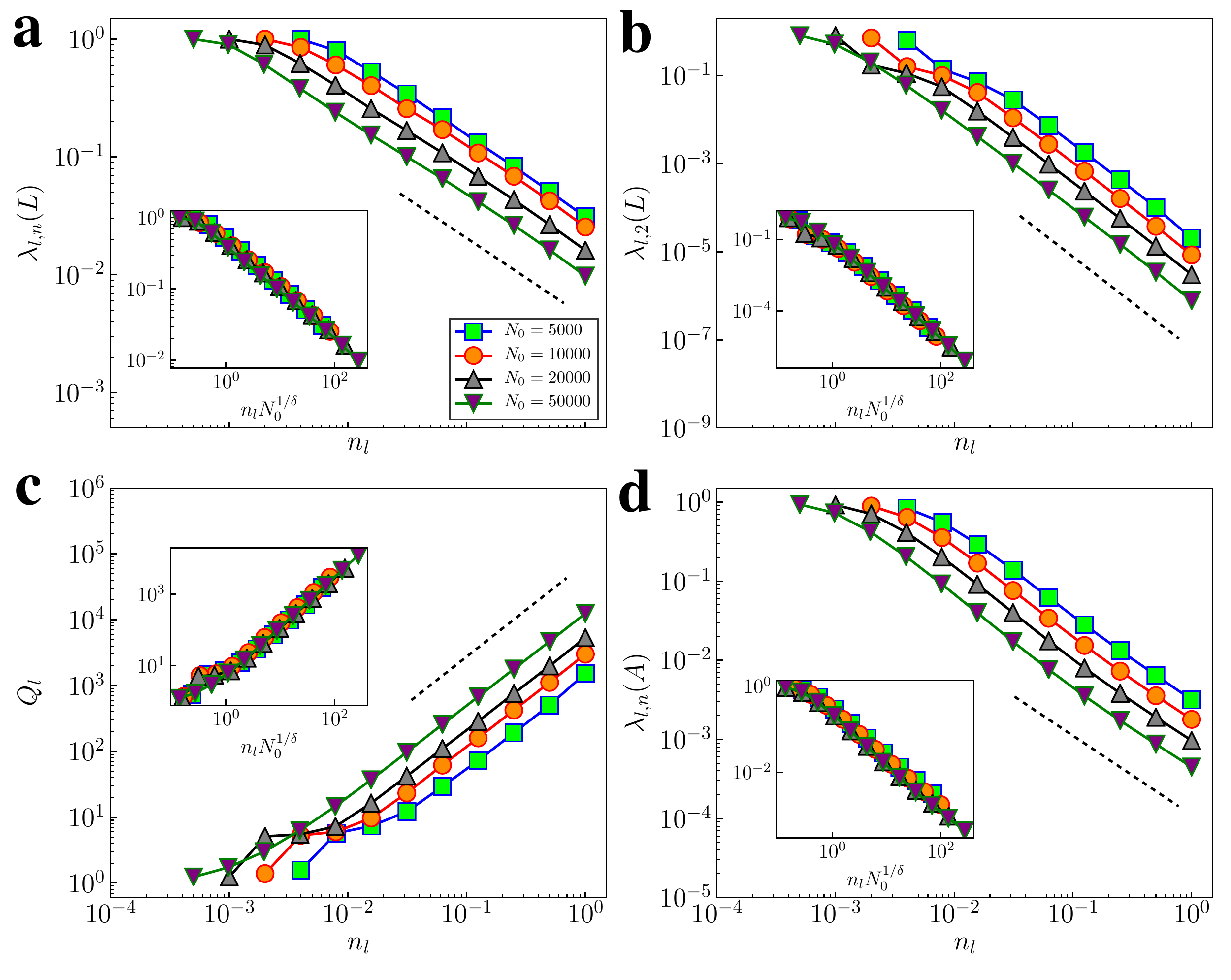}
\caption {FSS analysis of $\mathbb{S}^1$-network dynamic observables along the GR flow. The main figures show each observable as a function of the variable $n_l$ in the process of GR transformation, and the inset shows their scaling functions related to the variable $n_{l} N_{0}^{1/\delta}$. Key parameters of the $\mathbb{S}^1$ network are $\nu = 2.5$ and $\sigma = 1.5$, respectively; the scaling exponent corresponding to four observables is $\delta = 2 \pm 0.1$. For all observables, the black dashed-lines predicted their power-law behavior. The average hidden degree $\langle\kappa\rangle_0 \approx 6$ and the expected average degree $\langle k \rangle_0 \approx 6$ for all initial networks. All the results are averaged over 10 independent realizations.}
\label{Fig:2}
\end{figure}

Now, consider several other observables that represent global properties of networks. One is the normalized maximum eigenvalue of the Laplace matrix,
\begin{equation}\label{Eq:3}
\lambda_{l, n}(L)=\frac{\Lambda_{l, n}(L)}{N_{l}},
\end{equation}
where $\Lambda_{l, n}(L)$ is the maximum eigenvalue of the Laplace matrix $L$ of the renormalized network $G_l$, and $N_l$ is the maximum value that $\Lambda_{l, n}(L)$ may take. For a connected network with at least one edge, it always satisfies $\Lambda_{l, n}(L) \geqslant K_l + 1$, where the equality holds if and only if $K_l = N_l -1$~\cite{Brouwer2012}. The first normalized nonzero eigenvalue of the Laplace matrix is
\begin{equation}\label{Eq:4}
\lambda_{l, 2}(L)=\frac{\Lambda_{l, 2}(L)}{N_{l}},
\end{equation}
where $\Lambda_{l, 2}(L)$ is the first nonzero eigenvalue of the Laplace matrix $L$ of the renormalized network $G_l$, and $N_l$ is the maximum value that $\Lambda_{l, 2}(L)$ may take. To some extent, the functional properties of a network can be optimized by increasing the value of $\Lambda_{l,2}(L)$. For instance, maximizing $\Lambda_{l,2}(L)$ can maximize the rate of convergence to the network homogeneous state for undirected networks~\cite{Nishikawa2017}.

Also, the $\mathbb{S}^1$ model parameters can be divided into three regions according to the diffusion time $1/\Lambda_{l, 2}(L)$ as discussed before~\cite{Garcia2018}. In the following, consider the ratio of the maximum eigenvalue of the Laplace matrix to the first nonzero eigenvalue~\cite{Barahona2002,Nishikawa2003,Donetti2005,Nishikawa2006a,Nishikawa2006b,Brede2010},
\begin{equation}\label{Eq:5}
Q_{l}=\frac{\lambda_{l, n}(L)}{\lambda_{l, 2}(L)},
\end{equation}
which is related to the synchronizability~\cite{Barahona2002,Shi2013} and stability~\cite{Garcia2018} of the network synchronization process.

Lastly, the spectral properties of the network adjacency matrix determine the behavior of many dynamic processes,
among which the largest eigenvalue of the adjacency matrix is an important one, and its normalized
result is given by
\begin{equation}\label{Eq:6}
\lambda_{l, n}(A)=\frac{\Lambda_{l, n}(A)}{N_{l} - 1},
\end{equation}
where $\Lambda_{l,n}(A)$ is the largest eigenvalue of the adjacency matrix $A$ of $G_l$, and $N_l - 1$ is the maximum value that $\Lambda_{l,n}(A)$ may take. Recently, the relationship between the maximum eigenvalue $\Lambda_{l,n}(A)$ and two network subgraphs is revealed for a large number of synthetic and real networks~\cite{Castellano2017}. It also shows~\cite{Restrepo2005} the impact of $\Lambda_{l,n}(A)$ on two highly correlated dynamical models, one is epidemic spreading with threshold $\lambda_{c}=1 / \Lambda_{l, n}(A)$ and the other is synchronization of Kuramoto oscillators with threshold $\zeta_{c}=\zeta_{0} / \Lambda_{l, n}(A)$. In this context, the variables in Eqs.~\eqref{Eq:3}--\eqref{Eq:6} have a critical effect on the dynamical behavior of the network. For this reason, they are called dynamic observables of networks. Their FSS is shown in Fig.~\ref{Fig:2}.

Next, along the directions of the GR flows, FSS analysis is performed on eight network observables in regions \textbf{I} (small-world phase) and \textbf{II} (non-small-world phase), respectively. Specifically, the dependence of these observables on $n_l$ is investigated for each layer of the renormalized network. The results indicate that these observables can be represented by a scaling function with $n_{l} N_{0}^{1 /\delta}$ as the variable. More precisely, any observable $\mathcal{X}$ approximately satisfies
\begin{equation}\label{Eq:7}
\mathcal{X} = f\left(n_{l} N_{0}^{1 / \delta}\right),
\end{equation}
where $f(\cdot)$ is a function depending on the initial network size and specific transformation used.

\section*{Results}
\textbf{FSS of GR flows in synthetic networks.} Figure~\ref{Fig:1} shows the dependence of $k_{l, \max }$, $\langle k\rangle_{l, n}$, $\langle c\rangle_{l}$ and $\langle\ell\rangle_{l}$ on $n_l$ for the $\mathbb{S}^1$ synthetic network. The inset shows each observable as a function of $n_{l} N_{0}^{1 / \delta}$. The results show that the observable curves of networks with different sizes largely overlap. Specifically, in the small-world phase with $(\nu, \sigma) = (2.5, 1.5)$, the scaling exponent $\delta \approx 2$ (see Fig.~1), while in the non-small-world phase with $(\nu, \sigma) = (3.5, 2.5)$, the scaling exponent $\delta \approx 1$ (see Supplementary Information Fig.~S4). This appears to be true also when the values of $(\nu, \sigma)$ are taken elsewhere in each phase (see Supplementary Information Table~S1 for details). Interestingly, for $k_{l, \max }$ and $\langle k\rangle_{l, n}$, the results shown in Fig.~\ref{Fig:1} demonstrate that they are both approximately obey a power-law relationship with $n_l$, where the black dashed-lines predict their power-law behavior.

In the following, the power-law behaviors of these two observables are further discussed. Simulation results show that the maximum degree $K_l$ of the renormalized network $G_l$ and the maximum degree $K_{l - 1}$ of $G_{l - 1}$ are related as
\begin{equation}\label{Eq:8}
K_{l}=s^{\varepsilon} K_{l-1}=\cdots=s^{l \varepsilon} K_{0}.
\end{equation}
Since $n_{l} = N_{l} / N_{0} = s^{-l}$, it follows that
\begin{align}\label{Eq:9}
k_{l, \max } &=\frac{K_{l}}{N_{l}-1} \approx \frac{s^{l \varepsilon} K_{0}}{N_{l}}=\frac{n_{l}^{-\varepsilon} K_{0}}{n_{l} N_{0}} \notag\\
&=\frac{n_{l}^{-(\varepsilon+1)}K_{0}}{N_{0}} \notag \\
& \approx k_{0, \max } n_{l}^{-(\varepsilon+1)}.
\end{align}
The above equation indicates that $k_{l, \max }$ approximately follows a power-law relation, $k_{l, \max } \sim n_{l}^{-\beta}$, where $\beta = \varepsilon + 1$. For the observed average degree $\langle k\rangle_{l}$ of the renormalized network $G_l$, when $\nu - 1 < 2 \sigma$, it was shown~\cite{Garcia2018} that the average degree approximately satisfies an exponential relation along the GR flow, $\langle k\rangle_{l}=s^{\alpha}\langle k\rangle_{l-1}$, which yields
\begin{equation}\label{Eq:10}
\langle k\rangle_{l}=s^{\alpha}\langle k\rangle_{l-1}=\cdots=s^{l \alpha}\langle k\rangle_{0},
\end{equation}
consequently,
\begin{align}\label{Eq:11}
\langle k\rangle_{l, n} &=\frac{\langle k\rangle_{l}}{N_{l}-1} \approx \frac{s^{l \alpha}\langle k\rangle_{0}}{N_{l}}=\frac{n_{l}^{-\alpha}\langle k\rangle_{0}}{n_{l} N_{0}} \notag \\
&=\frac{n_{l}^{-(\alpha+1)}\langle k\rangle_{0}}{N_{0}} \notag \\
& \approx\langle k\rangle_{0, n} n_{l}^{-(\alpha+1)}.
\end{align}
The above equation shows that, when $\nu - 1 < 2 \sigma$, $\langle k\rangle_{l}$ approximately obeys a power-law relation, $\langle k\rangle_{l, n} \sim n_{l}^{-\eta}$, with $\eta = \alpha+1$. For $\nu - 1 > 2 \sigma$, through simulations it is found that the average degree $\langle k\rangle_{l}$ still approximately satisfies $\langle k\rangle_{l}=s^{\alpha}\langle k\rangle_{l-1}$, leading to $\langle k\rangle_{l, n} \sim n_{l}^{-\eta}$, with $\eta = \alpha+1$. The values of $\beta$ and $\eta$ are given in the Supplementary Information Table~S2.

\begin{figure}[!h]
\centering
\includegraphics[angle=0,width=17.6cm,height=5.28cm]{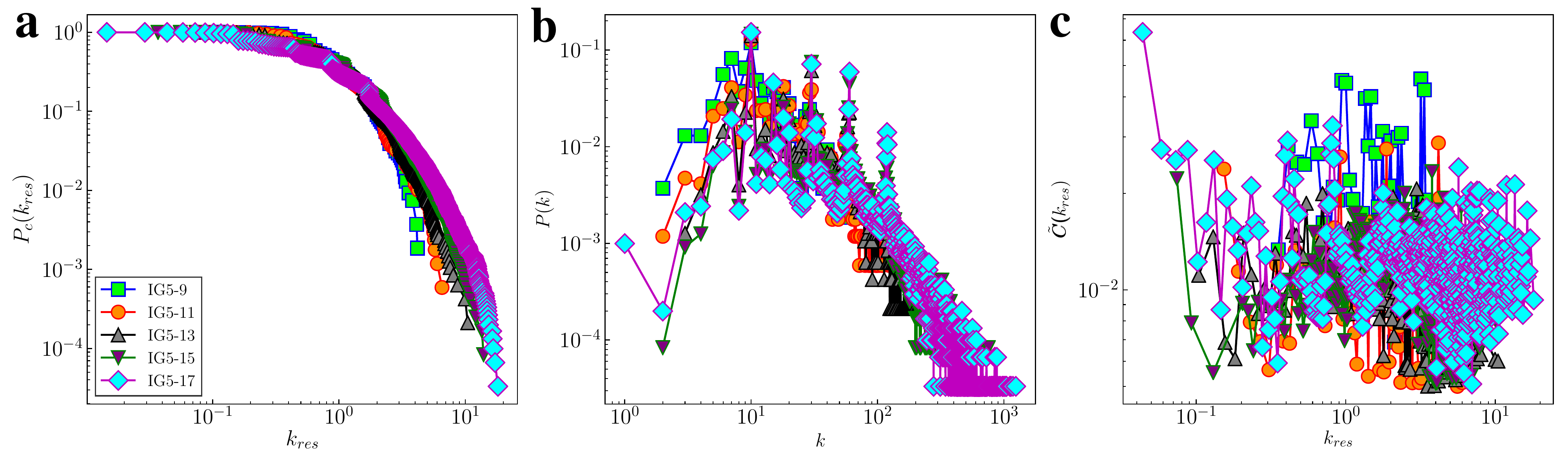}
\caption {Self-similarity of IG5 evolutionary networks. a The complementary cumulative distribution function (CCDF) $P_c$ of node rescaled degrees $k_{res} = k/\langle k\rangle$. b The probability distribution function (PDF) $P(k)$ of the degree of nodes. c The degree-dependent clustering coefficient $\tilde C(k_{res})$ of node rescaled degrees $k_{res}$. Topological characteristics of this series of evolutionary networks are given in Supplementary Information Table S3.}
\label{Fig:3}
\end{figure}

\begin{figure}[!h]
\centering
\includegraphics[angle=0,width=17.5cm,height=14cm]{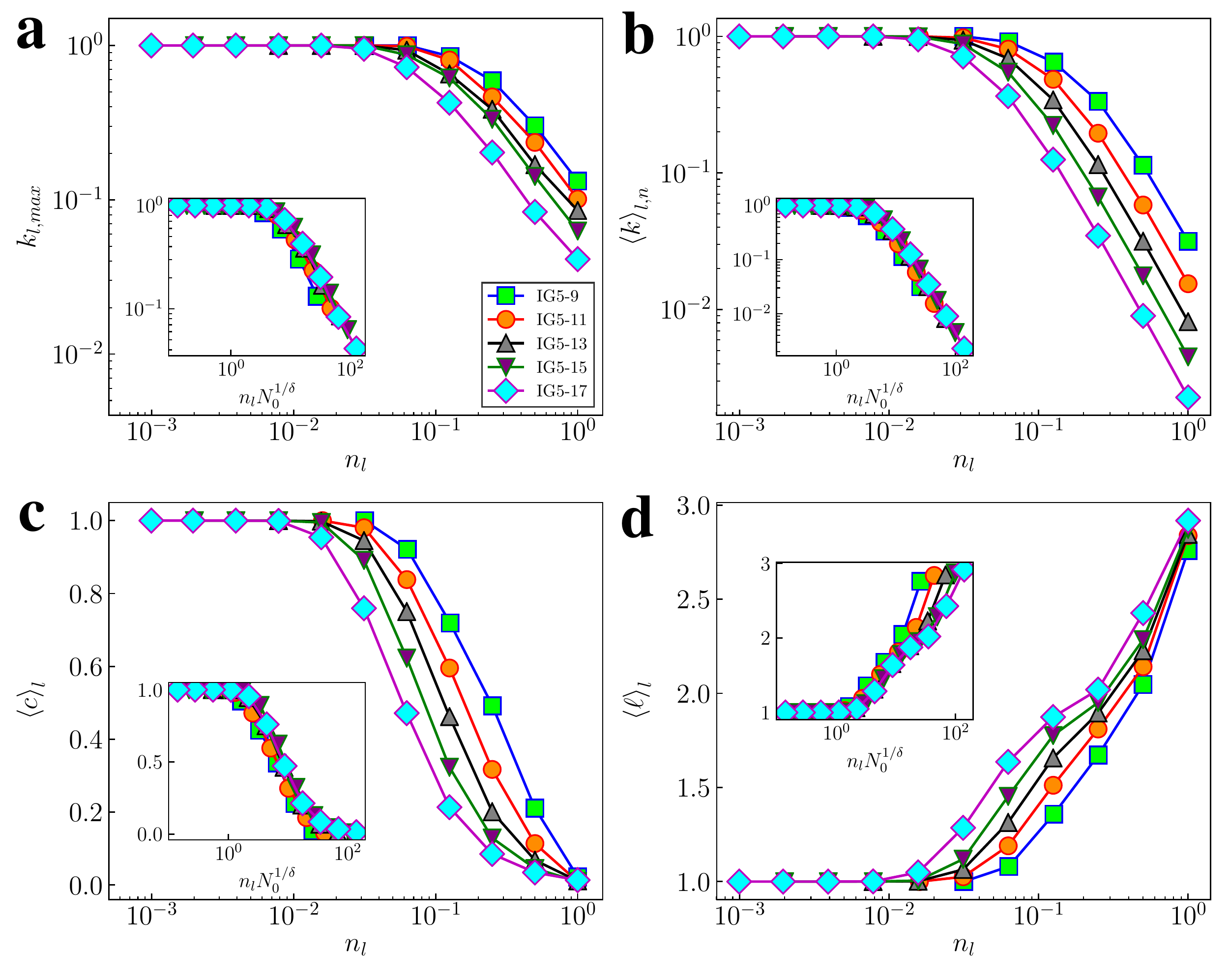}
\caption {FSS analysis of IG5 evolutionary networks' topological observables along the GR flow. The main figures show each observable as a function of the variable $n_l$ in the process of GR transformation, and the inset shows their scaling functions related to the variable $n_{l} N_{0}^{1/\delta}$. These networks belong to phase \textbf{I}, and the scaling exponent corresponding to four observables is $\delta = 2 \pm 0.1$.}
\label{Fig:4}
\end{figure}

Finally, consider the dependence of several dynamic observables on $n_l$ (see Eqs.~\eqref{Eq:3}--\eqref{Eq:6}), which depend on the spectral properties of the Laplace matrix and the adjacency matrix of the network, as shown in Fig.~\ref{Fig:2}. To a certain extent, these observables are able to reflect some dynamical properties of a network, such as synchronization stability, diffusion time, and synchronization threshold of Kuramoto oscillator parameters. The inset of Fig.~\ref{Fig:2} shows the dependence of the observables on $n_{l} N_{0}^{1 / \delta}$ for different sizes of networks, which is similar to the phenomenon presented in Fig.~\ref{Fig:1}. More importantly, the exponent $\delta$ is also consistent with that in Fig.~\ref{Fig:1}, namely, in the small-world phase, $\delta \approx 2$, and in the non-small-world phase, $\delta \approx 1$ (see Supplementary Information Fig.~S5). The black dashed-line predicts the power-law behavior of each observable along the GR flow. The corresponding power-law exponent values are listed in the Supplementary Information Table~S2. It is worth noting that these observables approximately obey power-law curves along the GR flows, providing important guidance for predicting the structural and dynamical properties of large-scale networks. For instance, one can use the eigenvalue ratio $Q_l$ of a renormalized smaller-size network to estimate the synchronizability of the initial large-scale network. To some extent, it can also be used to eliminate various consequences caused by the high complexity of large-scale networks. Furthermore, the results also show that for higher-dimensional embedded networks, by embedding a network into a $D$-dimensional ($D \geqslant 2$) space, the value of the exponent $\delta$ is consistent with that of the $\mathbb{S}^1$ model (see Supplementary Information Figs.~S8-S11).

\begin{figure}[!h]
\centering
\includegraphics[angle=0,width=17.5cm,height=14cm]{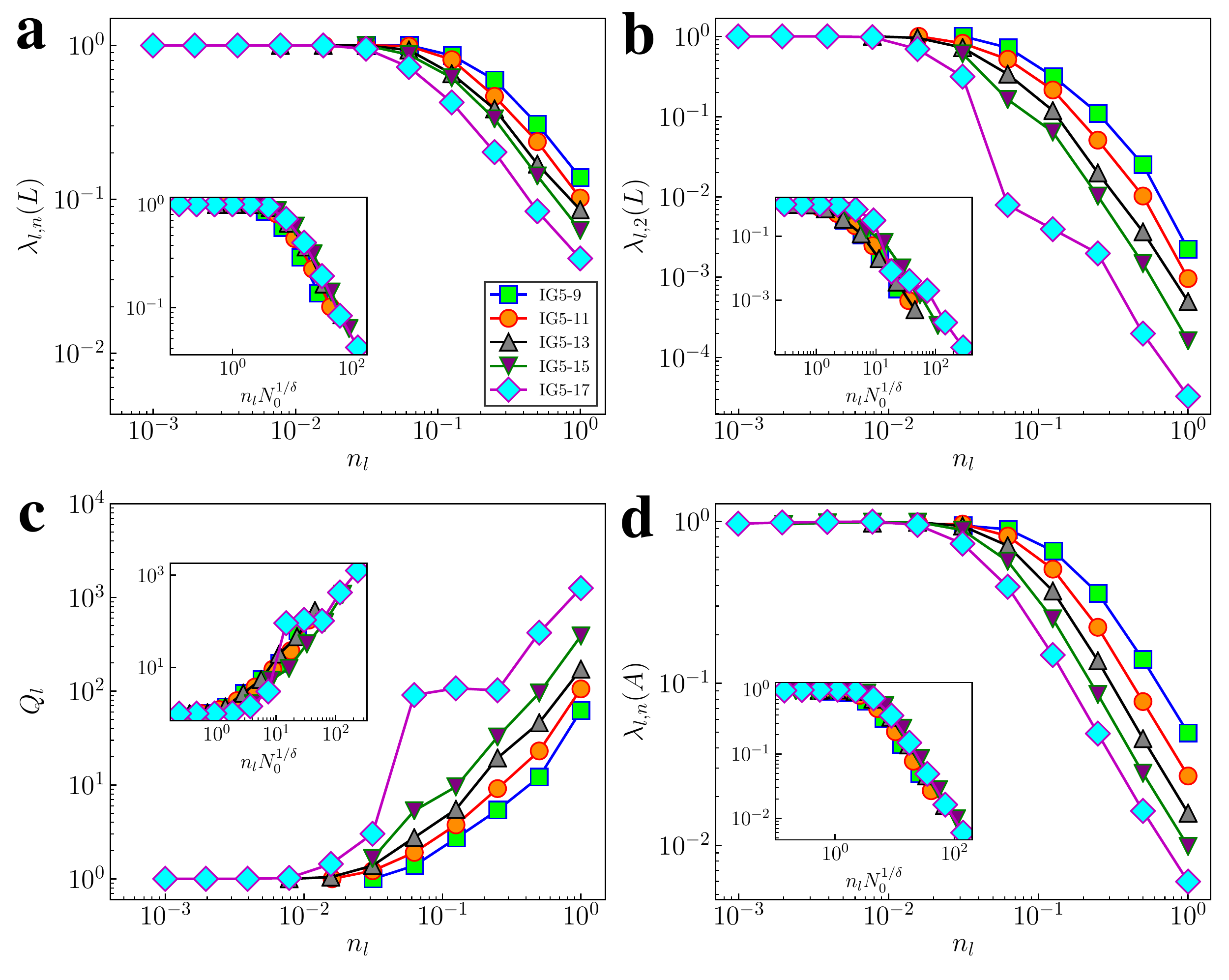}
\caption {FSS analysis of IG5 evolutionary networks' dynamic observables along the GR flow. The main figures show each observable as a function of the variable $n_l$ in the process of GR transformation, and the inset shows their scaling functions related to the variable $n_{l} N_{0}^{1/\delta}$. The scaling exponent corresponding to four observables is $\delta = 2 \pm 0.1$.}
\label{Fig:5}
\end{figure}


\textbf{FSS of GR flows in real evolutionary networks.} 
As a common practice, it is necessary to verify the universality of the above conclusions with real networks, mostly small-world networks. Obviously, this brings up the problem that a single network has only one initial size $N_0$. To address this issue, some evolutionary network systems are employed, each of which eventually leads to a series of networks of different sizes over time, and these networks of the same type with different sizes are all approximately in the same phase on the $(\nu, \sigma)$ plane. Ten types of real evolutionary networks~\cite{Rossi2015,Leskovec2007,Ripeanu2002,Leskovec2005,Newman2001} (including 39 networks) are investigated and the parameter $\sigma$ value of each network is inferred according to an existing method~\cite{Garcia2019}. The results show that these networks belong to small-world networks, that is, in phase \textbf{I} or \textbf{III}.

These networks are then embedded into different $\mathbb{S}^1$ networks and then the geometric renormalization transformation is performed. It is found that the eight observables curves of the same type but different sizes of the evolutionary networks overlap roughly, and the exponent $\delta$ is consistent with that of the synthetic network (small-world phase), i.e., $\delta \approx 2$. Take the IG5 evolutionary network~\cite{Rossi2015} as an example. It has five different system sizes at five-time points (see Supplementary Information Table S3), and these five networks all belong to phase \textbf{I}. Fig.~\ref{Fig:3} shows the complementary cumulative distribution function, probability distribution function and degree-dependent clustering coefficient versus node degrees. Examining these characteristics, they are self-similar. The FFS analysis is performed along the GR flow, and the results suggest that there is a phenomenon similar to that of synthetic networks, that is, the curves of each observable almost overlap under the scaling function with $n_l N_0^{1/\delta}$ as the variable, and the exponent $\delta \approx 2$ (see Fig.~\ref{Fig:4} and Fig.~\ref{Fig:5}). The other nine types of evolutionary networks produced similar results (see Supplementary Information Figs.~S12-S38), further confirming that the exponent $\delta$ can be used as a effective measure for dividing small-world and non-small-world networks.

\section*{Discussion}
In conclusion, the scaling behaviors of structural and dynamical observables of $\mathbb{S}^1$ synthetic and real evolutionary networks have been systematically investigated along the GR flows. According to the structural properties of the $\mathbb{S}^1$ model, the network evolutionary phase can be divided into three regions. Some networks with different structural parameters are generated in each region, with finite-size scaling analysis on their structural and dynamical observables. The results show that these observables can be characterized by a certain scaling function with $n_{l} N_{0}^{1 / \delta}$ as the variable. More importantly, the critical exponent $\delta$ is found to be independent of these observables but dependant only on the small-world properties of the network. More precisely, networks located in the small-world phase region all have exponent $\delta = 2$, while those located in the non-small-world phase region and in their critical regions all have $\delta = 1$. This implies that the $\mathbb{S}^1$ model can be divided into two universal types according to the value of the exponent $\delta$.

Inspired by the findings on model $\mathbb{S}^1$, some real evolutionary network systems are considered, typically in the small-world phase, which also follows the finite-size scaling found in model $\mathbb{S}^1$. Specifically, for each system, there are different sizes at different time points. By embedding these networks of different sizes into the $\mathbb{S}^1$ network, it was found that they all approximate the same phase on the $(\nu, \sigma)$ plane, and these networks are self-similar. Thus, this investigation continues to study the scaling laws of some real evolutionary systems in the same way as the synthetic networks. The findings further suggest that the $\mathbb{S}^1$ model can provide more evidence for predicting the structural and dynamical behavior of those real networks. On the other hand, the results of this paper can provide some guidance for studying the structural and functional characteristics of large-scale networks. For instance, for an evolutionary network, often with a relatively small initial size, will eventually evolve to a large-scale system, so that it is difficult to obtain its structural and functional characteristics via computer simulation. While the scaling law found in this paper makes it possible to predict the characteristics of large-scale networks from small-size networks. Furthermore, the new results also show that the GR transformation may lead to significant changes of some properties of the network, which can be captured by the scaling function under the finite-size scaling analysis.

\bibliographystyle{unsrt}
\bibliography{manuscript}

\section*{Acknowledgments}
We sincerely acknowledge G. Garc\'ia-P\'erez, M. Bogu\~n\'a, and M. \'A. Serrano for sharing the codes of the geometric renormalization of general networks, and their suggestions have played an important role in improving the quality of the paper. This work was supported by the National Natural Science Foundation of China under Grant Nos.~61991412 and 61873318, the Frontier Research Funds of Applied Foundation of Wuhan under Grant No.~2019010701011421, and the Program for HUST Academic Frontier Youth Team under Grant No.~2018QYTD07.

\section*{ Author contributions}
All authors conceived the work, analyzed results, and wrote the manuscript. D. Chen wrote Python implementations. 

\section*{Competing interests}
The authors declare that they have no competing interests. 

\section*{Data and materials availability}
The data that support the plots within this paper and other findings of this study are available from the corresponding author upon request.

\end{document}